\documentclass[12pt,preprint]{aastex}
\usepackage{graphicx}
\usepackage{epstopdf}
\usepackage{amsmath}

\begin{document}

\title{Evolution of Low-Mass X-ray Binaries: the Effect of Donor Evaporation}

\author{Kun Jia$^{1,2}$ and Xiang-Dong Li$^{1,2}$}

\affil{$^{1}$Department of Astronomy, Nanjing University, Nanjing 210046, China}

\affil{$^{2}$Key Laboratory of Modern Astronomy and Astrophysics (Nanjing University), Ministry of Education, Nanjing 210046, China}

\affil{$^{}$lixd@nju.edu.cn}

\begin{abstract}
Millisecond pulsars (MSPs) are thought to originate from low-mass X-ray binaries (LMXBs).
The discovery of eclipsing radio MSPs including redbacks and black widows indicates that evaporation of the donor star by the MSP's irradiation takes place during the LMXB evolution.
In this work, we investigate the effect of donor evaporation on the secular evolution of LMXBs, considering different evaporation efficiencies and related angular momentum loss. We find that for widening LMXBs, the donor star leaves a less massive white dwarf than without evaporation; for contracting systems, evaporation can speed up the evolution, resulting in dynamically unstable mass transfer and possibly  the formation of isolated MSPs.
\end{abstract}


\section{INTRODUCTION}
In neutron star low-mass X-ray binaries (NS LMXBs), mass transfer proceeds through Roche-lobe overflow (RLOF) of the low-mass donor star. The accreting material, preserving its own angular momentum (AM), forms a disk around the NS. During this disk accretion process, the NSs can acquire both mass and AM while releasing gravitational energy in the form of high-energy radiation. In this regard, slowly rotating NSs are recycled into millisecond radio pulsars (MSPs) by mass transfer \citep[][for reivews]{Bhattacharya1991PhR,Tauris2006csxs}. Notably, several transitional pulsars were recently observed to swing between accretion-powered X-ray sources and rotation-powered radio pulsars, lending strong support to the evolutionary link between MSPs and LMXBs \citep[e.g.,][]{Archibald2009Sci,Papitto2013Natur,Roy2014ATel}.

Theoretically, the evolution of LMXBs and their connection with MSPs have been extensively investigated in various aspects. However, some key processes remain uncertain. One of them is the amount and the mechanism of mass and angular momentum loss (AML) during the mass transfer episode.
The recent discovery of  eclipsing radio MSPs called redbacks  \citep[][for a review]{Roberts2013IAUS} suggests new possible paths in the evolution of LMXBs. These objects share  similarities with the well-known black widow MSPs \citep{Fruchter1988Natur}, although they have more massive companions and wider orbits. It is widely accepted that in black widow MSPs the companion stars are bloated by the high-energy radiation and/or particles from the MSPs \citep{Phinney1988Natur,vandenHeuvel1988Natur,Ruderman1989ApJa}, and similar processes also occur in redback MSPs. Their positions in the companion mass - orbital period diagram demonstrate that they are unlikely to be the end products of LMXB evolution, but emerge due to temporary cessation of mass transfer in LMXBs\footnote{Most of the redback companions are non-degenerate normal stars. They are nearly RL-filling systems and may switch to the LMXB phase as seen in the
three transitional MSPs.}. The evolutionary relationship between redback and black widow MSPs is still in debated, and several possible formation mechanisms have been proposed \citep{Chen2013ApJ,Benvenuto2014ApJ,Smedley2015MNRAS,Jia2015ApJ}. Nevertheless, all these models converge on introducing the donor evaporation boosted by the energetic MSP's irradiation, and the evaporative winds would certainly affect the secular evolution of LMXBs and their end products. This evaporation mechanism may also be closed related to the formation of  isolated MSPs.

In this work, we investigate the LMXB evolution taking into account the influence of donor evaporation, and pay particular attention to the possible final products of different evolutionary sequences. The rest of this paper is organized as follows. In Section 2 we introduce the physical considerations and binary evolution model. Our results are presented in Section 3, and discussed in Section 4. We conclude in Section 5.

\section{MODEL}

In this study, we calculate the LMXB evolution with the Modules for Experiments in Stellar Astrophysics (MESA) code \citep{Paxton2011ApJS,Paxton2013ApJS,Paxton2015ApJS}. We configure and modify the code to construct the binary models we expect to investigate. The basic physical considerations of the model are same as in \cite{Jia2015ApJ} and briefly described as below.

We start the evolution from an incipient LMXB system composed of a NS (defined as the primary) of mass  $M_{1}$ and a zero-age main-sequence (ZAMS) companion (defined as the secondary/donor) of mass $M_{2}$. We take solar chemical abundance for the companion star.
Its RL radius is evaluated with the \cite{Eggleton1983ApJ} formula
\begin{equation}
\frac{R_\mathrm{L,2}}{a}=\dfrac{0.49q^{2/3}}{0.6q^{2/3}+\ln(1+q^{1/3})},
\end{equation}
where $q=M_{2}/M_{1}$ is the mass ratio and $a$ is the orbital separation.
We adopt the \citet{Ritter1988AA} scheme in MESA to calculate the mass transfer rate via RLOF, and assume that a fraction $\beta$ of the transferred matter is accreted onto the NS while the rest is ejected out of the system from the NS in the form of isotropic winds. We fix $\beta=0.5$, and limit the NS accretion rate is by the Eddington rate, i.e.,
\begin{equation}
\dot{M}_1=\min(-\beta\dot{M}_{2},\dot{M}_{\rm Edd}),
\end{equation}
with
\begin{equation}
\dot{M}_{\rm Edd}=3.6\times10^{-8}\left(\frac{M_{1}}{1.4~M_{\odot}}\right)\left(\frac{0.1}{GM_{1}/R_{1}c^{2}}\right)\left(\frac
{1.7}{1+X}\right)M_{\odot}\rm yr^{-1}.
\end{equation}
Here $-\dot{M}_2$ is the mass transfer rate, $X$ is the hydrogen abundance, and $R_{1}$ is the NS radius taken to be $10^{6}~\rm cm$.

When the mass transfer rate drops below a critical value the accretion disk becomes thermally unstable due to the unbalance between viscous heating and cooling in the disk \citep{Meyer1982AA,Smak1982AcA}. We adopt the the criterion for the critical mass transfer rate in \cite{Lasota2008AA} for an X-ray irradiated accretion disk,
\begin{equation}
\dot{M}_{\rm cr}=9.5\times10^{14}C_{-3}^{0.36}\alpha_{0.1}^{0.04+0.01\mathrm{log}C_{-3}}R_{\rm d,10}^{2.39-0.10\mathrm{log}C_{-3}}(M_{1}/M_\sun)^{-0.64+0.08\mathrm{log}C_{-3}}~\mathrm{gs^{-1}},
\end{equation}
where the constant $C=10^{-3}C_{-3}$ accounts for the irradiation efficiency \citep{Dubus1999MNRAS}, $\alpha=0.1\alpha_{0.1}$ is the viscosity parameter, and $R_{\rm d}=10^{10}R_{\rm d,10}$  cm is the outer radius of the accretion disk given by \citep{Paczynski1977ApJ}
\begin{equation}
R_{\mathrm{d}}=\dfrac{0.60a}{1+q}.
\end{equation}
An unstable disk experiences limit cycles displaying short outbursts with long quiescent intervals. We assume no accretion during quiescence and  accretion during outbursts takes place at a rate of $\dot{M}_1=\min(-\beta\dot{M}_{2}/d,\dot{M}_{\rm Edd})$, were $d$ is the duty cycle and we adopt a fixed $d = 0.01$ \citep{King2003MNRAS} in our calculation.

Besides the isotropic winds during the mass transfer, we introduce evaporative winds  from the secondary star when it suffers the MSP's irradiation  \citep{Ruderman1989ApJa}. The corresponding mass loss rate is given by \citep{vandenHeuvel1988Natur},
\begin{equation}
\dot{M}_{2,\rm evap}=-\dfrac{f}{2v_{2,\rm esc}^{2}}L_{\rm p}\left(\dfrac{R_{2}}{a}\right)^{2},
\end{equation}
where $f$ is the efficiency of the MSP's irradiation, $v_{2,\rm esc}$ the surface escape velocity of the secondary star, $L_{\rm p}$ the spin-down luminosity of the MSP, and ${R_{2}}$ the radius of the secondary.
To calculate the spin-down luminosity $L_{\rm p}=4\pi^{2}I\dot{P}/P^{3}$ caused by magnetic dipole radiation   \citep{Shapiro1983bhwd}, we adopt typical values for the parameters of MSPs, i.e., the moment of inertia $I=10^{45}\,\rm gcm^{2}$, the initial spin period $P=3$ ms, and the initial spin derivative $\dot{P}=1.0\times10^{-20}\,\rm ss^{-1}$. Same as in \citet{Jia2015ApJ}, we consider  evaporation to take place once the following two conditions are satisfied: (1) the NS has  been fully recycled with accreted mass more than $0.1~M_{\odot}$, and (2) the accretion disk becomes thermally unstable, so the radio activity can turn on during the quiescent episodes.

The total AML during the evolution is as follows,
\begin{equation}
\dot{J}=\dot{J}_\mathrm{GR}+\dot{J}_\mathrm{ML,NS}+\dot{J}_\mathrm{ML,evap}+\dot{J}_\mathrm{MB}.
\end{equation}
The first term $\dot{J}_\mathrm{GR}$ on the right hand side of Eq.~(7) is due to gravitational radiation (GR) \citep{Landau1975ctf},
\begin{equation}
\dot{J}_\mathrm{GR}=-\dfrac{32}{5}\dfrac{G^{7/2}}{c^{5}}\dfrac{M_{1}^{2}M_{2}^{2}(M_{1}+M_{2})^
{1/2}}{a^{7/2}}\rm~,
\end{equation}
where $c$ is the speed of light and $G$ is the gravitational constant.
The second ($\dot{J}_\mathrm{ML,NS}$) and third ($\dot{J}_\mathrm{ML,evap}$) terms denote the AML related to the isotropic winds from the NS and  the evaporative winds from the secondary, respcetively.
Considering the uncertainty in how the evaporative winds leave the system,
here we adopt two possible modes: Mode A with the winds leaving from the surface of the secondary and Mode B  from the inner Lagrangian (L1) point. The related AML depends on the distance from the center of mass of the binary to the position where the winds leave the system, i.e.,
\begin{equation}
d_\mathrm{A}=\dfrac{a}{1+q},
\end{equation}
for Mode A, and
$$d_\mathrm{B}=\left\{
\begin{aligned}
& (0.5-0.227\mathrm{log}~q - \dfrac{q}{1+q})a, ~0.1 \leqslant q \leqslant 10 ~ \\
& \left[ 1-\left(\dfrac{q}{3(1+q)}\right)^{1/3}-\frac{1}{3}\left(\dfrac{q}{3(1+q)}\right)^{2/3}-\dfrac{q}{27(1+q)}\right] a, ~ q < 0.1 ~  .
\end{aligned}
\right. $$
for Mode B \citep{Kopal1959cbs,Frank2002apa}.
\noindent For a typical LMXB with $q\leqslant 1$, the specific AM with Mode B winds is less than with Mode A winds (in the extreme case of $q=1$, Mode B winds take no systematic AM away).

The last term is due to magnetic braking (MB), for which we adopt the standard formula given by \citep{Verbunt1981AA,Rappaport1983ApJ},
\begin{equation}
\dot{J}_\mathrm{MB}=-3.8\times10^{-30}M_{2}R_{2}^{4}\omega^{3}\ {\rm
dyn\,cm},
\end{equation}
where $\omega$ is the angular velocity of the binary.

\section{RESULTS}

\begin{figure}
\centering
\plotone{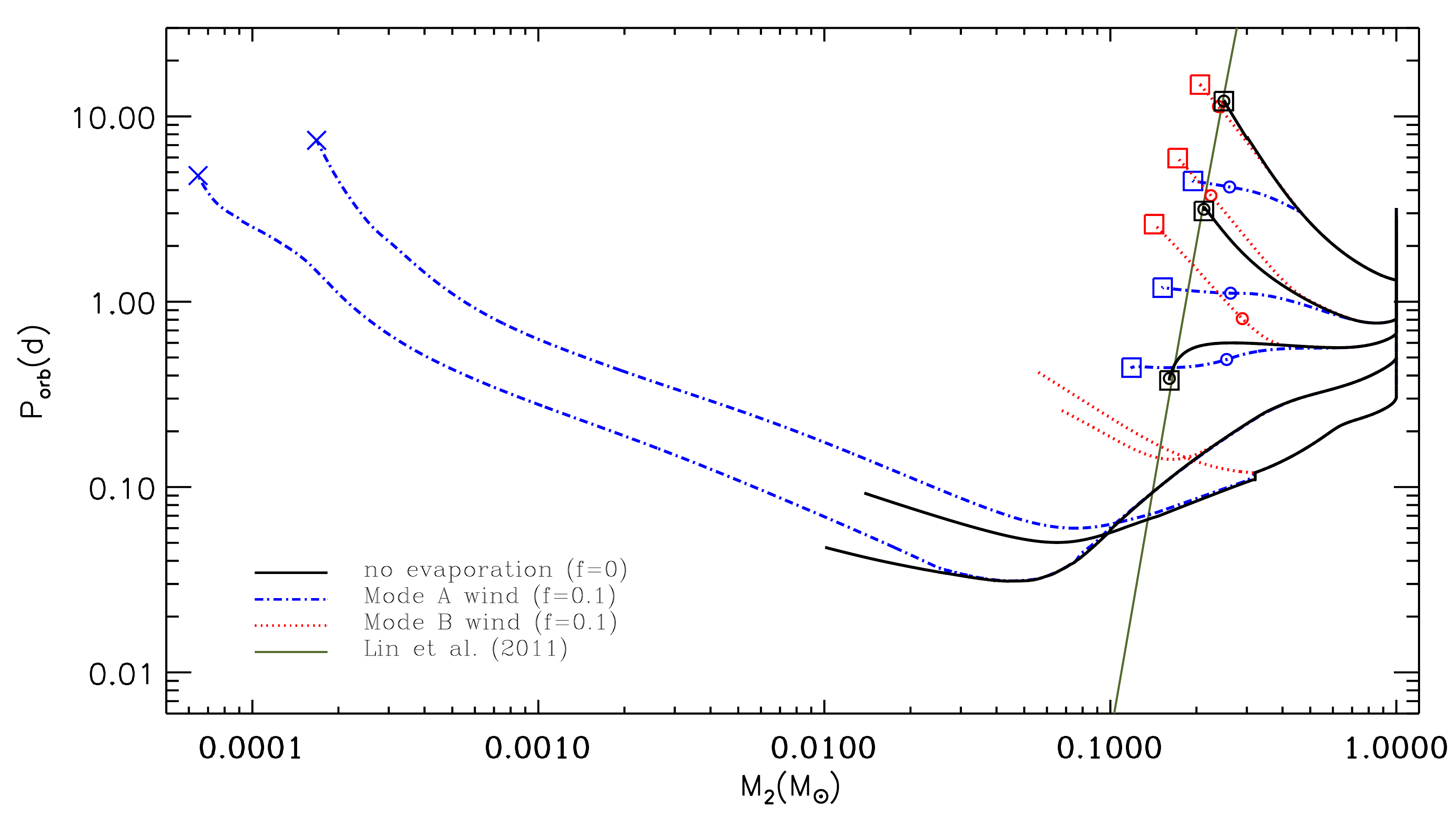}
\caption{Comparison of the evolutionary tracks of LMXBs with and without evaporation in the $M_{\rm 2} - P_{\rm orb}$ plane. The solid, dot-dashed and dotted lines correspond to the standard model ($f=0$), the models of Modes A and B evaporation ($f=0.1$), respectively. The open circles and squares indicate the birth of proto-WDs and the beginning of their cooling phase, respectively.
All the calculations are ended at Hubble age, except for the tracks ended with crosses which represent the start of runaway mass transfer. The initial secondary mass is 1 $M_{\odot}$, and the initial periods for each group are 3.2, 2.9, 2.75, 2.4, and 0.5 d.
The thin solid line represents the theoretical $M_{\rm WD} - P_{\rm orb}$ relation given by \cite{Lin2011ApJ} for Pop~I stars.
\label{figure1}}
\end{figure}

As demonstrated by previous studies, donor evaporation can influence both stellar and orbital evolution  \citep[e.g.,][]{Ergma2001MNRAS,vanHaaften2012AA,Benvenuto2012ApJ,Benvenuto2014ApJ,Chen2013ApJ,Smedley2015MNRAS,Jia2015ApJ}.
Figure~\ref{figure1} compares the evolutionary tracks with and without evaporation for LMXBs in the $M_{2}$ vs. $P_{\rm orb}$ plane. For each group of the selected binary sequences, the initial parameters (i.e., the orbital period $P_{\rm orb, i}$ and the secondary mass $M_{2, \rm i}$) are the same, but the evaporation modes are different: the solid, dot-dashed, and dotted lines correspond to no evaporation, Modes A and B evaporation with $f=0.1$, respectively. There exists a bifurcation period ($\sim 0.5-1$ d) in the LMXB
evolution \citep{Pylyser1988AA,Pylyser1989AA}. Binaries with $P_{\rm orb,i}$ below the bifurcation period evolve along the cataclysmic variable (CV)-like tracks with shrinking orbits caused by AML, while in binaries with $P_{\rm orb,i}$ above the bifurcation period, mass transfer is driven by the donor's nuclear expansion, and the evolution leads to the formation of a recycled NS and a white dwarf (WD), which is the degenerate core of the donor star  \citep[][for a recent review]{Deloye2008AIPC}.
Also plotted with the thin solid line is the theoretical white dwarf mass ($M_{\rm WD}$) - orbital period ($P_{\rm orb}$) relation for MSP binaries in \citet{Lin2011ApJ}. We can see that the evolution can be seriously altered under evaporation, depending on the donor mass, evaporation efficiency, and mass loss and AML efficiencies.
Generally, for orbit-widening tracks, evaporation causes the final binaries to deviate from the $M_{\rm WD}-P_{\rm orb}$ relation, and the secondary masses to be smaller; for shrinking tracks, evaporation can significantly affect the stellar evolutionary process and even induce a final unstable mass transfer (see discussion below).

In Fig.~\ref{figure2}, we plot the evolutionary tracks of the secondary star in the $\log g - T_{\rm eff}$ plane (where $g$ and $T_{\rm eff}$ are the surface gravity and effective temperature, respectively), which act as a diagnosis for the produced WDs. The left and right panels correspond to the cases without evaporation and with Mode A evaporation (with $f=0.1$), respectively.
To illustrate the nature and the evolutionary state of the secondary star, we use different colors to display the magnitude of the central degeneracy parameter $\psi_{\rm c}$.
The thick and thin lines represent the RLOF and RL-detachment stages, respectively.
In the widening systems, the secondary stars have an opportunity to evolve into the red-giant branch with a burning hydrogen shell and an expanding envelope. When the burning shell is running out, the $g$-value reaches a minimum. After that, the secondary star rapidly contracts and terminates RLOF. During the subsequent proto-WD phase, the surface temperature increases to a maximum along with contraction. Then the WD enteres the cooling track. To demonstrate the entire sequence we extend the calculation up to a time of 20 Gyr, and use a red star in each track to mark Hubble time.

In the following we discuss the widening and contracting evolutionary tracks separately.

\subsection{The $M_{\rm WD}$ - $P_{\rm orb}$ relation for widening binaries}

\begin{figure}
\centering
\plotone{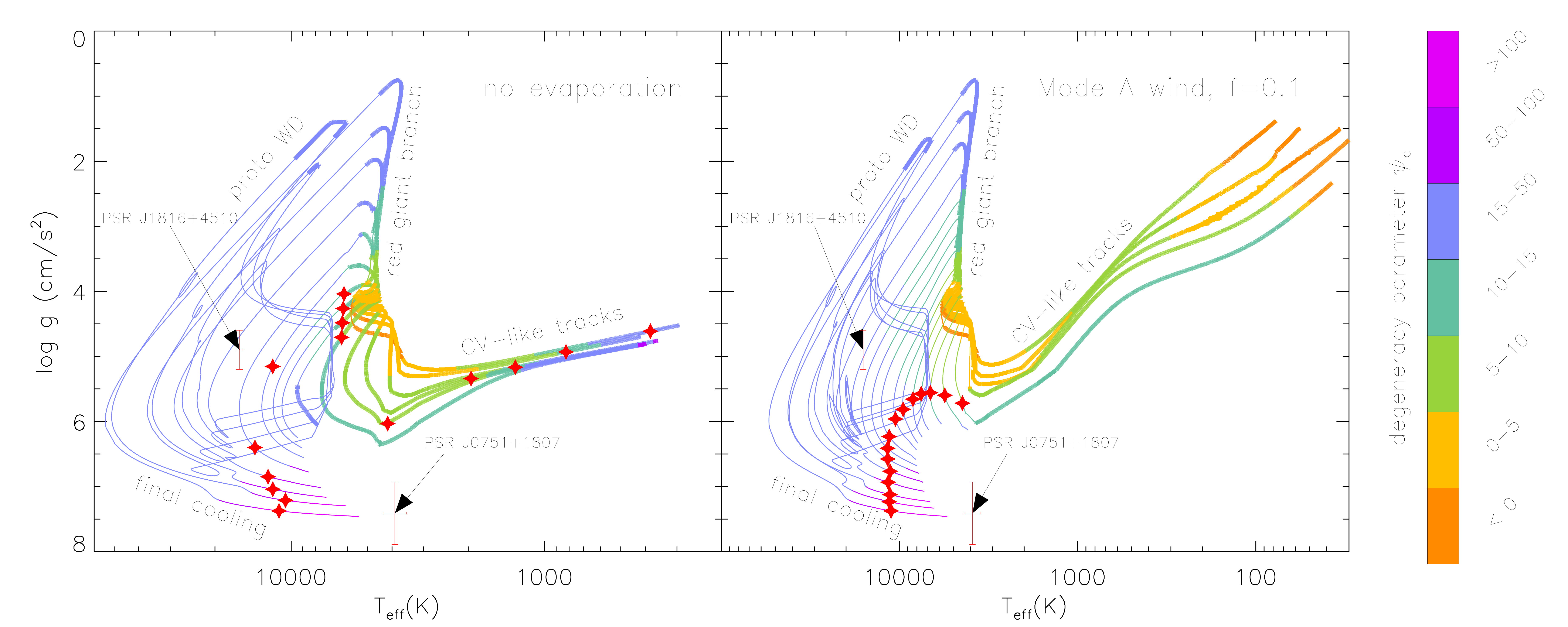}
\caption{Selected evolutionary sequences of the secondary star in the $\log~g - T_{\rm eff}$ plane. The left and right panels correspond to the cases of no evaporation and Mode A evaporation with $f=0.1$, respectively. The thick and thin lines represent the RLOF and RL-decouplement phases, respectively. The colors describe the magnitude of the central degeneracy parameter $\psi_{\rm c}$ of the secondary star. The red star symbol on each sequence marks Hubble time.
\label{figure2}}
\end{figure}

The $M_{\rm WD}$ - $P_{\rm orb}$ relation is an important expectation from the LMXBs evolution with initial orbital period longer than the bifurcation period,
which originates from the relation between the mass of the degenerate helium (He)-core and the radius of the red-giant donor \citep[e.g.,][]{Webbink1983ApJ,Taam1983}. At the end of the red-giant phase, the donor star starts to shrink, terminating mass transfer and finally leaving a WD remnant. In previous works the lower end of the $M_{\rm WD}$ - $P_{\rm orb}$ relation relies on whether the proto-WD can shrink to underfill its RL \citep{Smedley2014MNRAS,Jia2014ApJ,Istrate2014AAa}. However, this RL-detachment criterion may not work when RLOF is not the unique way of mass loss, since RL decoupling may take place due to evaporation-induced orbital expansion when the secondary is still on the main sequence. Figure~\ref{figure2} shows that the evolutionary products of the secondary stars transit continuously from degenerate WDs to non-degenerate stars. In fact, a
nuclear evolved star can be always contracting and become degenerate if there's enough time and no mass loss. The minimum $\sim0.13~M_{\odot}$ WD mass \citep{Tutukov1987SvAL} for single star evolution which results
from the Sch\"{o}nberg-Chandrasekhar limit \citep{SC1942ApJ} may not apply in the binary case.
In this sense, there would not be a lower mass limit for WDs if their progenitors are continuously stripped in a binary system. Therefore, to study the $M_{\rm WD}$ - $P_{\rm orb}$ relation we need a physical definition of WDs. Essentially, WDs are stellar objects supported by degeneracy pressure of electron gas. For degenerate matter, the Fermi energy $E_{F}$ is much larger than the thermal energy $kT$, and the degeneracy parameter $\psi \equiv E_{F}/kT$ is an useful indicator to describe the degree of degeneracy. In fact, ideally complete degeneracy exists only when $\psi\rightarrow\infty$ (or $T\rightarrow0$), and in real situations all WDs should be partially degenerate. In this work, we consider the central degeneracy $\psi_{\rm c}>10$ as a prerequisite for WDs. Additionally, WDs should be the remnants of giant stars and have entered the final cooling tracks. Finally,  if the WD is detached from its RL within Hubble time, then we regard that  a MSP-WD binary is formed.

In Fig.~\ref{figure1} we mark the proto-WD phase in each evolutionary track, between the minimum of $\log g$ at the end of giant phase and the maximum of $T_{\rm eff}$ at the beginning of the final cooling with open circles and crosses, respectively. Such a proto-WD phase is also shown in Fig.~\ref{figure2}. Note that  a proto-WD can still loss mass under evaporation. When it has entered the final cooling track, the evaporative winds are much reduced, and there's little evolution in the $M_{2}-P_{\rm orb}$ plane.

\begin{figure}
\centering
\plotone{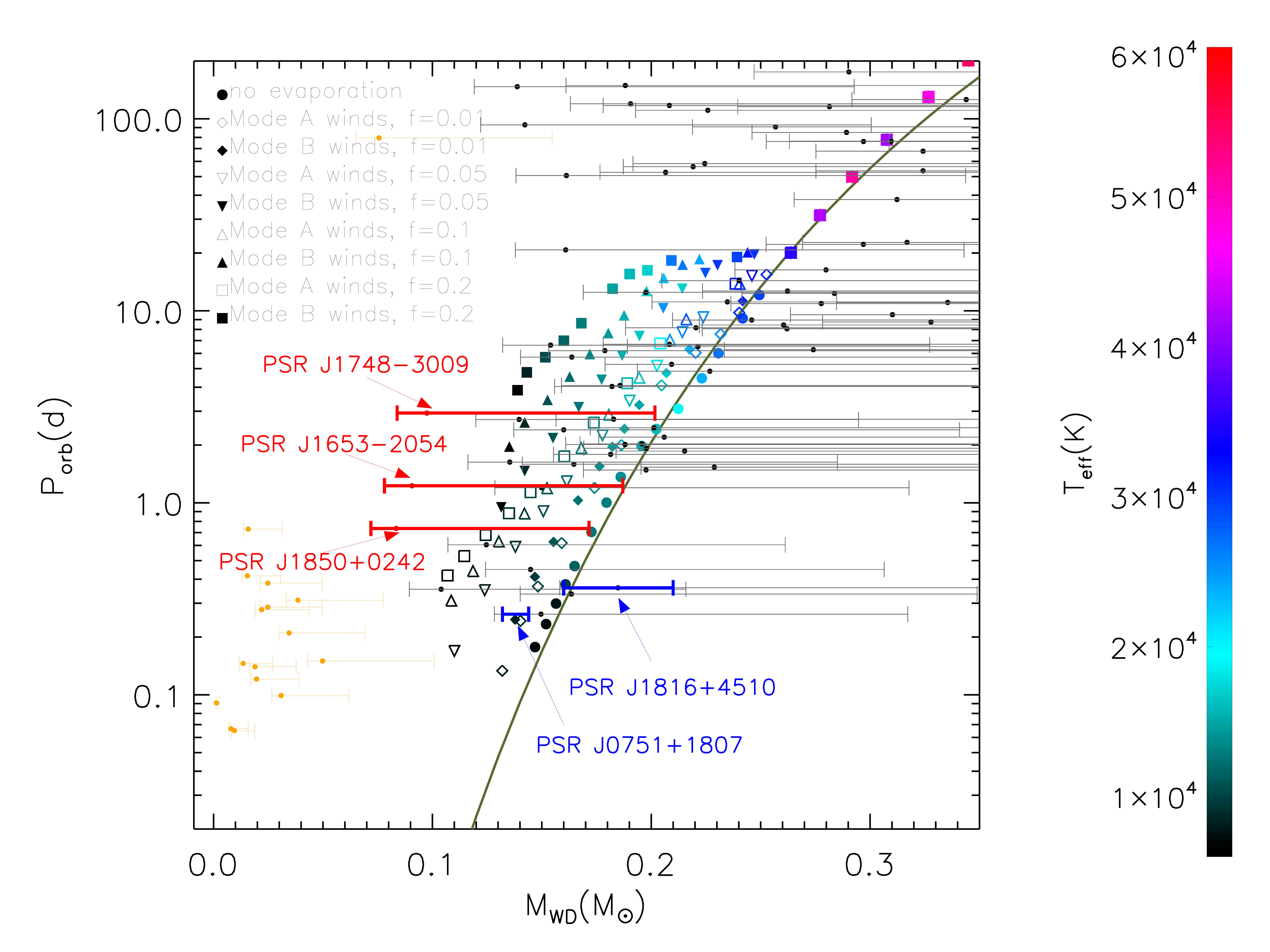}
\caption{The calculated $M_{\rm WD} - P_{\rm orb}$ relation based on different evaporation efficiencies and AML modes, depicted with different symbols. The colors represent the maximum surface temperature of the WD at the beginning of the cooling phase. The thin solid line shows the theoretical $M_{\rm WD} - P_{\rm orb}$ relation given by \cite{Lin2011ApJ} for Pop~I stars. The small black dots represent the Galactic binary MSPs in the field with WD companions and the orange ones with ultra-light (with mass $< 0.08~M_{\odot}$) companions, and the error bars correspond to orbital inclination from $90^{\circ}$ to $26^{\circ}$, covering $90\%$ probability the WD mass distribution. Some interesting systems PSRs J0751+1807, J1653-2054, J1748-3009, J1816+4510, and J1850+0242 are also indicated.
\label{figure3}}
\end{figure}

Based on the consideration above, we plot the calculated $M_{\rm WD}$ vs. $P_{\rm orb}$ under different evaporation efficiencies and AML modes in Fig.~\ref{figure3}. A remarkable feature is that the $M_{\rm WD}$ - $P_{\rm orb}$ relation beomes rather scattered under evaporation. For $P_{\rm orb}\lesssim 20\rm~d$, the values of $M_{\rm WD}$ are smaller than predicted by the traditional model shown with the solid black curve\footnote{For systems with $P_{\rm orb}\gtrsim 20\rm~d$, we don't introduce evaporation-induced mass loss into the calculation,  since these NSs cannot accrete $0.1~M_{\odot}$ to become MSPs in our calculation. This may be an over-simplified assumption since MSPs are also found in such wide systems \citep[see][for a discussion]{Tauris1996,Shao2012}.}. This is easy to understand, since the traditional $M_{\rm WD}$ - $P_{\rm orb}$ relation  reflects the correlation between the radius of the secondary star and the binary orbital period, but the WD progenitor may be underfilling its RL due to the orbital expansion under evaporation. So the traditional $M_{\rm WD}$ - $P_{\rm orb}$ relation represents a lower limit of the orbital period. Therefore, we consider that such  evaporation-induced deviation should be important for relatively compact ($P_{\rm orb}\lesssim 20\rm~d$) MSP-WD systems. To compare with observations, we plot in Fig.~\ref{figure3} the field binary MSPs with either WD or ultra-light (with mass $< 0.08~M_{\odot}$) companions (data taken from  the ATNF catalogue\footnote{http://www.atnf.csiro.au/people/pulsar/psrcat/}). For each pulsar, the error bars cover $90\%$ probability of the mass distribution for random orientation of the orbit. Although a large fraction of the MSPs seem to be compatible with the solid curve, there do exist some MSPs which are located to the left of the solid black curve, including PSRs J0751$+$1807, J1850$+$0242, J1653$-$2054 and J1748$-$3009. Considering the uncertainties in the WD masses, optical observations can help to reveal the nature of these systems \citep{vanKerkwijk2005ASPC}.

\subsection{Evaporation accelerated evolution and final runaway mass transfer}

A considerable fraction ($\sim 1/3$) of MSPs are isolated and their origin remains to be a mystery. A likely mechanism is ablation of the companion stars by the MSPs, as observed in black widow systems \citep{vandenHeuvel1988Natur}, but whether such ablation can disrupt the secondary stars within Hubble time is an open question \citep{Kulkarni1988ApJ,Ruderman1989ApJa,Stappers1996ApJ,Chen2013ApJ}. Essentially, the destiny of an ablated secondary depends on the competition between mass loss and cooling-induced contraction. If the mass loss timescale is shorter than the thermal timescale, the secondary could be completely ablated. Otherwise it may continue contracting, and finally leave a degenerate remnant.

\begin{figure}
\plottwo{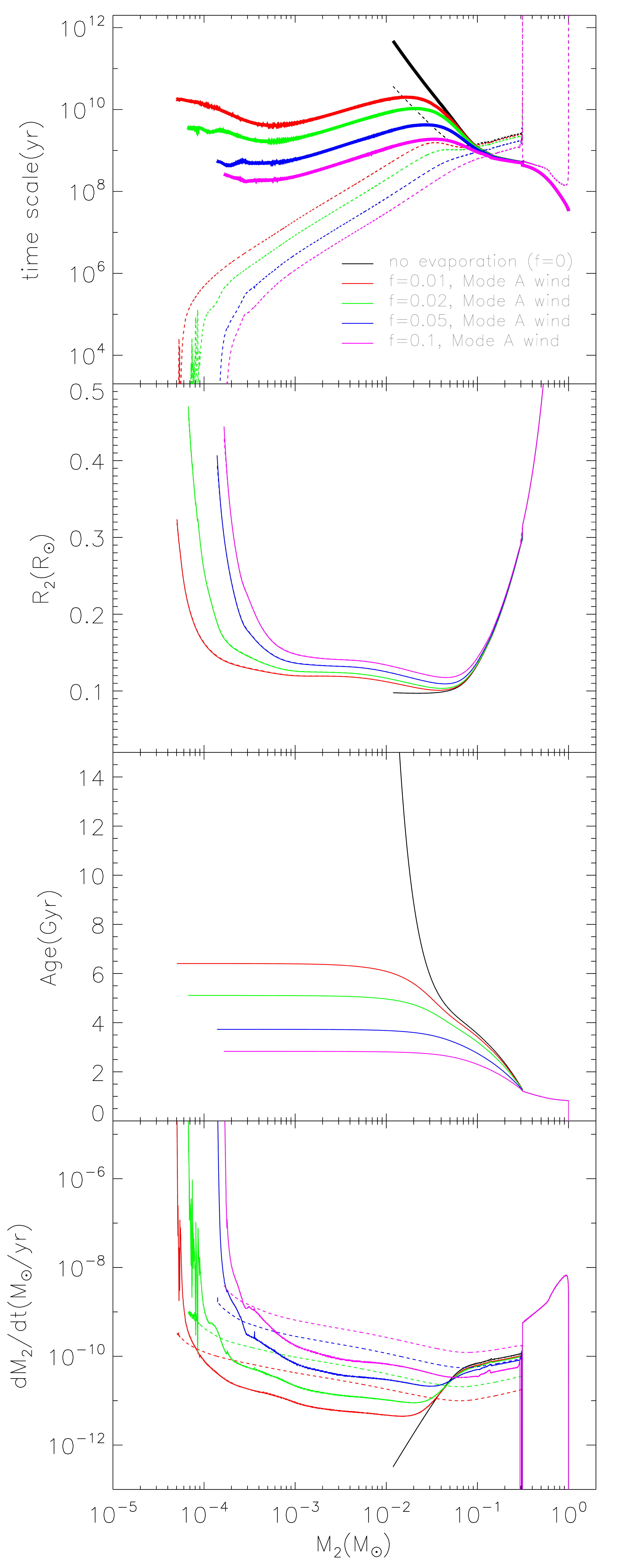}{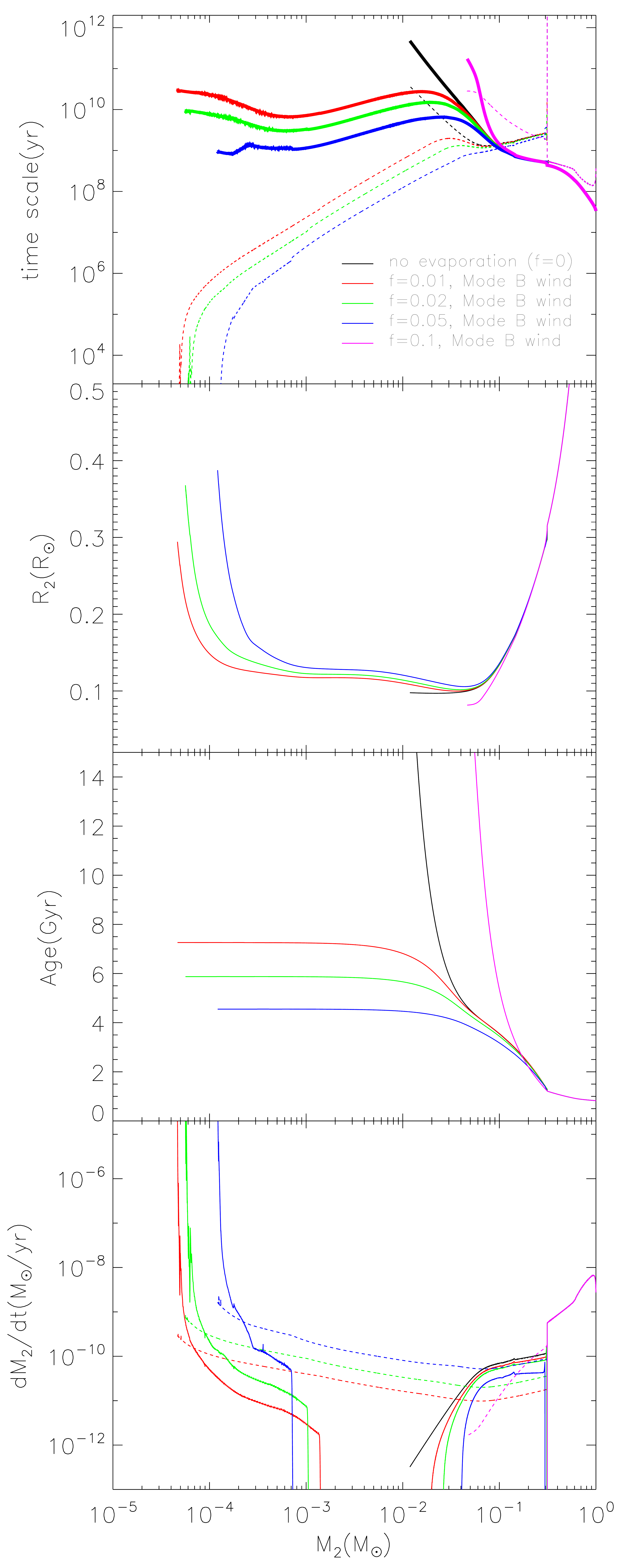}
\caption{Selected evolutionary tracks with initial secondary mass of $1.0~M_{\odot}$ and orbital period of $1.0\rm~d$ but different evaporation efficiencies and AML modes. Shown from top to bottom are the evolutionary timescale (thick lines for thermal timescale and thin dashed lines for mass loss timescale), radius, age, and mass loss rate of the secondary (thick lines for RLOF and thin dashed lines for evaporative mass loss), respectively.
\label{figure4}}
\end{figure}

Our calculations show that the evaporative winds can affect not only the orbital evolution, but also the thermal evolution of the secondary. In Fig.~\ref{figure4}, we plot a series of typical CV-like evolutionary tracks with the same initial binary parameters ($M_2=1.0~M_{\odot}$ and $P_{\rm orb}=1.0\rm~d$) but different evaporation efficiencies. The left panels are for the case of Mode A winds.
When the secondary is stripped to a mass below $\sim~0.1~M_{\odot}$, it becomes fully convective, and the mass loss timescale is shorter than the thermal timescale, so the star expands with mass loss. The radius-evolution plot shows that higher wind efficiency causes more rapid expansion and higher mass transfer rate. Mass stripping finally transforms the star into a sub-stellar object. When the secondary mass drops below $\sim2\times10^{-4}~M_{\odot}$, its self gravity cannot constrain the envelope, leading to a rapid rise of the mass transfer rate. At this time, the mass loss rate can be above $10^{-5}~M_{\odot}{\rm yr}^{-1}$ with typical duration $<100$ yr.

The right panels in Fig.~4 are for Mode B winds and show something different. With less efficient AML, Mode B winds lead to a more expanded binary orbit and the secondary star tends to underfill its RL. This results in a temporary suspension of RLOF as shown in the mass transfer rate plot. In the case of $f=0.1$ the stellar evolution does not finally lead to unstable mass transfer, but is dominated by cooling and contracting accompanied with weakened evaporative winds. Both the mass loss timescale and the thermal timescale exceed Hubble time, so the system would be observed as a black widow all the way with increasing orbital period. For smaller $f$, the secondary is able to re-fill its RL and lead to runaway mass transfer when its mass drops below $\sim 10^{-4}~M_{\odot}$.

\begin{figure}
\centering
{\includegraphics[width=\textwidth,height=\textheight,keepaspectratio]{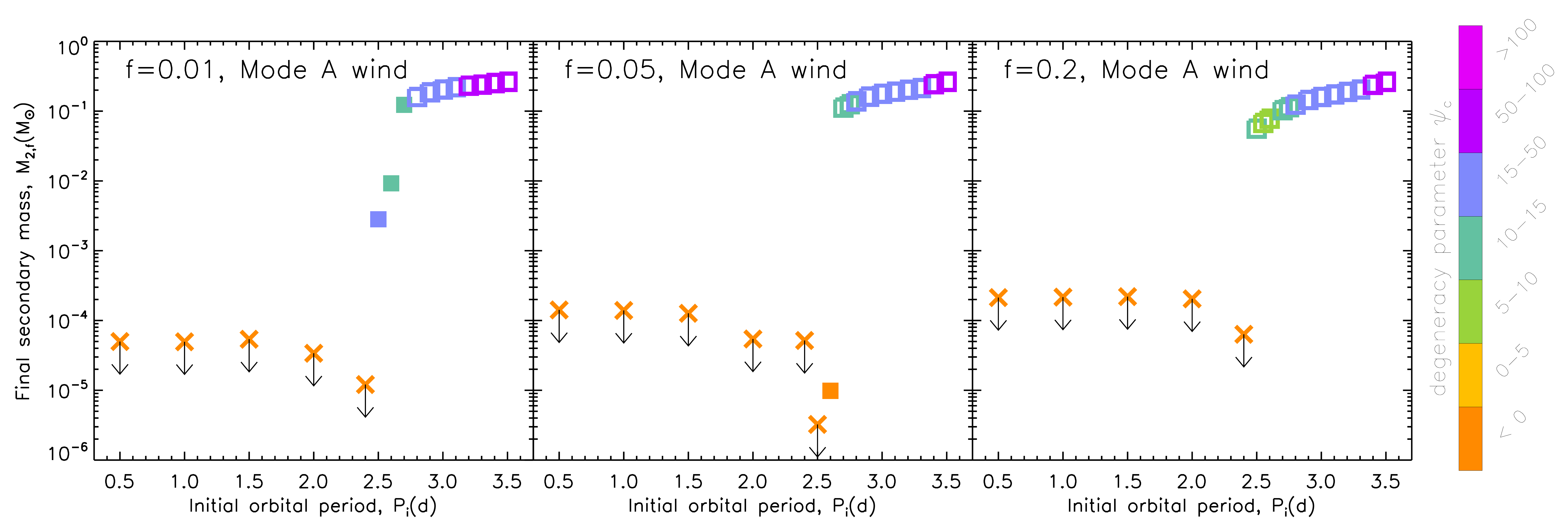}}
\centering
{\includegraphics[width=\textwidth,height=\textheight,keepaspectratio]{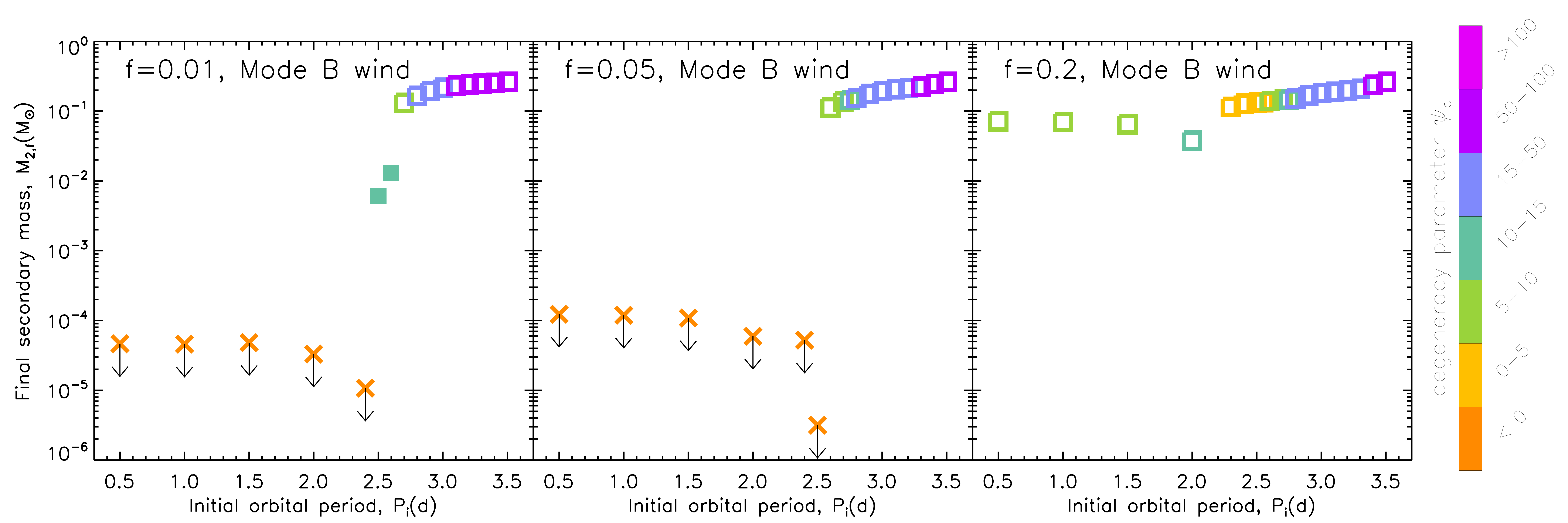}}
\caption{Final secondary mass as a function of the initial orbital period under various kinds of evaporative winds. The squares represent the secondary masses at Hubble age, with filled and open ones corresponding to RL-filling and under-filling systems respectively. The crosses represent the beginning of runaway mass transfer. The colors show the magnitude of the central degeneracy parameter $\psi_{\rm c}$ of the secondary stars.
\label{figure5}}
\end{figure}

In Fig.~\ref{figure5} we plot the final secondary mass as a function of the initial orbital period, with different evaporation efficiencies and wind modes. The squares represent the binaries at the age of Hubble time, with the filled and open ones indicating RL-filling and under-filling systems respectively; the crosses represent the state with dynamically unstable mass transfer. To demonstrate the nature of the secondaries, we use different colors to depict their central degeneracy parameter $\psi_{\rm c}$.
We can see that the evolution diverges at a modified bifurcation period $\sim2.5$ d. Below this period, the evolution follows the contracting CV-like tracks and leads to runaway RLOF except for the case of $f$=0.2 with Mode B winds.

The evolution following the runaway mass transfer is still uncertain. As suggested by \cite{Stevens1992MNRAS}, it may lead to the disruption of the companion star, forming a massive disk around the accretor. In our work, this runaway mass transfer occurs with a planet-mass donor which is different from the case of a (low-mass) stellar-mass donor in \cite{Stevens1992MNRAS}.
Therefore, the MSP companion might be disrupted in our case, leaving an isolated MSP. Previous works have shown that the timescale for CV-like evolution generally exceeds Hubble time. However, when considering the evaporative winds, the evolution can be significantly speeded up, suggesting that evaporation-driven mass transfer may serve as a possible way to form isolated MSPs.

\section{DISCUSSION}

\subsection{Comparison with observations of MSP-He WD binaries}

\begin{figure}
\centering
\plotone{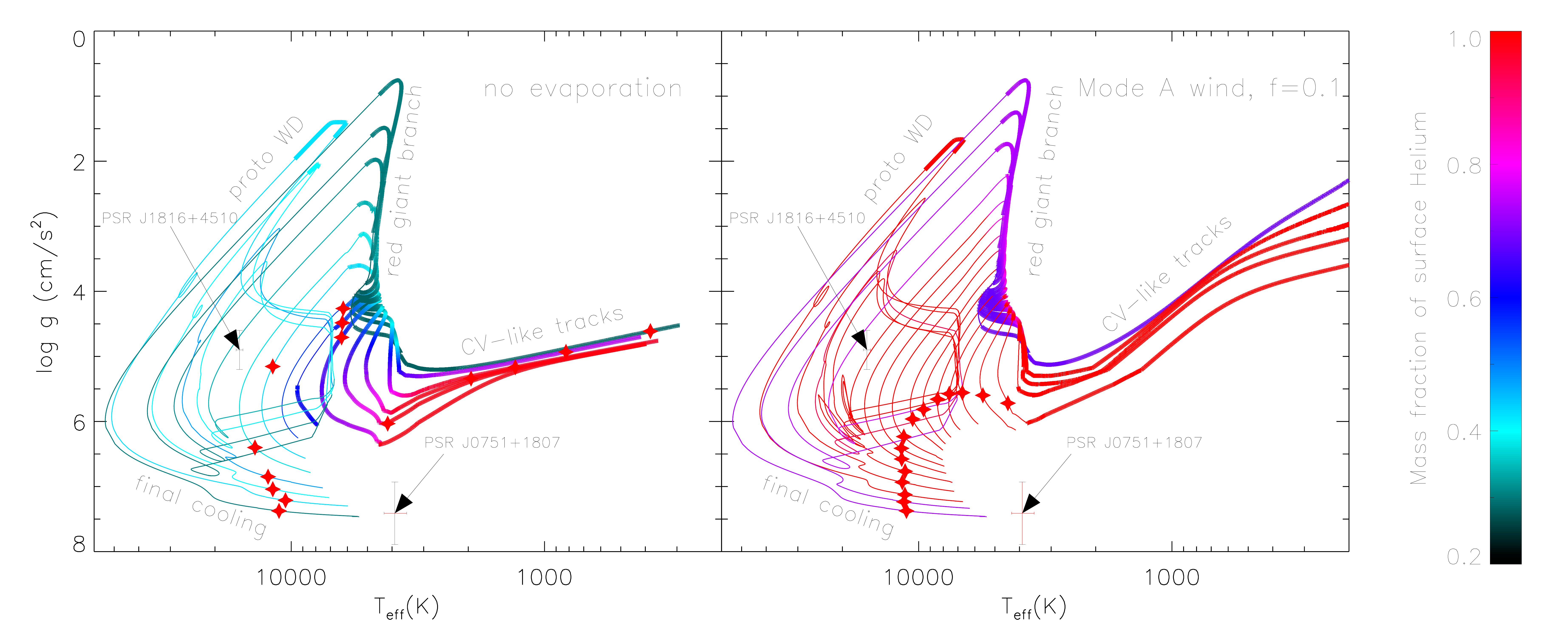}
\caption{Similar as Fig.~\ref{figure2} but with the colors indicating the the surface He mass fraction in the secondary stars.
\label{figure6}}
\end{figure}

As we have already shown,  evaporative winds can cause the final $M_{\rm WD}$ - $P_{\rm orb}$ distribution to significantly deviate from the traditional relation for MSPs. Besides, such winds can alter the surface chemical compositions of the secondary  in the cooling phase. In Fig.~\ref{figure6} we show how the surface He fraction changes along the evolutionary tracks in Fig.~\ref{figure2}. In the standard evolution,  He WDs with  mass $\sim 0.17-0.22~M_{\odot}$ are expected to preserve thick hydrogen-rich envelope,  avoiding hydrogen shell flashes \citep{Serenelli2001MNRAS,Panei2007MNRAS,Kilic2010ApJ}. However, Fig.~\ref{figure6} clearly shows hydrogen depletion accompanied with mass transfer under evaporation, thus hydrogen-depleted low-mass WDs may suggest possible wind ablation history. Additionally, hydrogen depleting can speed up  WD cooling, which may provide useful information about the binary history when compared with the MSP's spin-down \citep{Ergma2001MNRAS}.

Figure~\ref{figure3} shows that there are a few MSP-He WD systems located to the left the theoretical $M_{\rm WD}$ - $P_{\rm orb}$ relation. Among them PSR J0751$+$1807 (hereafter J0751) has the most accurately measured WD mass of $0.138(\pm0.006)~M_{\odot}$\footnote{David Nice, private communication}. However, its 6.31 hr orbital period is too wide compared with predicted by the $M_{\rm WD}-P_{\rm orb}$ relation. Optical observation suggested a hydrogen-depleted envelope of the WD with a pure He atmosphere (or a He atmosphere with some hydrogen mixed in) \citep{Bassa2006AAa}, which makes J0751 distinctive from other low-mass WDs which possess thick hydrogen-rich envelopes. These characteristics are compatible with the evaporation scenario. Moreover, to explain the discrepancy between the WD's cooling age and the spin down age of J0751, \cite{Ergma2001MNRAS} proposed that irradiation-driven mass loss could exhaust the hydrogen envelope and accelerate the cooling process.

Besides J0751, PSR J1816+4510 (hereafter J1816) is another outlier in which wind ablation is taking place, as it is a redback system \citep{Kaplan2012ApJ}. J1816 is also unique among the MSP-He WD binaries. The position of J1816's companion in the $\log~g$ - $T_{\rm~eff}$ plane (see Fig.~\ref{figure2}) indicates that it is actually a proto-WD on its cooling way \citep{Kaplan2013ApJ}. Thus, this individual source provides a good opportunity to study WD cooling under evaporation. Its secular orbital evolution is valuable for understanding the AML process associated with the evaporative winds.

The small number statistics and the uncertainties in both the measured parameters of MSP binaries and theoretical modeling prevent a reliable constraint on the evaporation process. However, some general tendencies can been seen in Fig.~3. With $f<0.2$ the evolutionary tracks with both Mode A and B winds can cover the distribution of the systems (with $P_{\rm orb}<20$ d) deviated from the $M_{\rm WD}-P_{\rm orb}$ relation. Larger values of $f$ lead to less massive $M_{\rm WD}$, and with the same $f$-value, $M_{\rm WD}$ is smaller in the case of Mode B winds than in the Mode A wind case.

\subsection{From black widows to isolated MSPs}

Our calculations suggest that CV-like evolution might disrupt the secondary or transform it into a planet-like object within Hubble time for both evaporation models. Observationally there are 38 MSPs among the total 42 pulsars that are accompanied with an ultra-light companion, and most of these MSPs have been identified as black widows\footnote{See also the catalogue at https://apatruno.wordpress.com/about/millisecond-pulsar-catalogue/.}.

The ultra-light companions of black widow MSPs will eventually either become crystallized (or even degenerate) or disrupted, depending on the competition between mass loss and cooling process. Recently \cite{Valsecchi2015ApJ} investigated the RLOF process of hot Jupiters with MESA, taking into account the effects of tides, irradiation, evaporated stellar winds, MB and so on. They found that after RLOF the remnant planets are left with rocky cores (hot Neptune or super-Earth), depending on the mass transfer process and the planetary core mass. In this work, the planet-like companions formed through CV-like evolution would not have the opportunity to form a rocky core in the center. In the right panel of Fig.~\ref{figure2}, we can see that the central degeneracy of the secondary star gradually decreases with the expansion of the radius. Figure~\ref{figure5} indicates that a considerable part of the black widow companions would be disrupted, and a few within a narrow range of the initial orbital period could be crystallized and form semi-degenerate stars with $\psi_{\rm c}<10$, but we caution that this result is sensitive to the wind modes.

Observationally the known MSP-planet systems are very rare. PSRs B1257$+$12 \citep{Wolszczan1992Natur} and J1719$-$1438 \citep{Bailes2011Sci} are the only two field MSPs with planets. PSR B1257+12 is in a multiple system possessing two planets about three times massive of Earth in 67 and 98 d orbits respectively, and a lunar mass object in a 25 d orbit. PSR J1719$-$1438 is a MSP  with a Jupiter-mass planet. Its 2.2 hr orbital period implies an ultra-compact X-ray binary origin  \citep{Bailes2011Sci,Benvenuto2012ApJ,vanHaaften2012AA}.
The population synthesis study by \cite{vanHaaften2013AA} also suggested that the predicted number of old UCXBs seems to match that of isolated MSPs. To investigate the possible products of UCXBs we calculate the evolution of an
UCXB consisting of a 1.3 $M_{\odot}$ NS and a 0.6 $M_{\odot}$ He star with chemical compositions $Y=0.98$ and $Z=0.02$. The initial orbital period is set to be 40 min. We adopted default wind mass loss rate prescription \citep{Nugis2000} for He stars in MESA. Figure 7 compares the evolutionary tracks with ($f=0.1$) and without evaporation considered. The orbital period first decays driven by GR. With growing mass-loss rate the orbital period reaches a minimum ($\sim 10$ min). After that the orbit starts to expand and the mass transfer rate declines all the way. When the orbital period is above $\sim 20$ min, the mass transfer rate drops below the critical value to maintain a stable disk \citep{Lasota2008AA}, and at this time evaporation starts to work, which accelerates RLOF \citep[see also][]{Heinke2013ApJ}. This increased mass transfer together with evaporation-induced mass loss can significant speed up the UCXB evolution, and it is possible for the MSP to completely ablate its companion within Hubble time.

Besides the MSP-planet systems, PSR B1937+21 is a MSP surrounded by an asteroid belt \citep{Shannon2013ApJ}. The planets or the asteroid belt around PSRs B1257+12 and B1937+21 were probably formed from a debris disk, but the origin of such a debris disk is quite uncertain \citep[][for a review]{Podsiadlowski1993ASPC}. Among the proposed models, our results are compatible with the disrupted companion model by \cite{Stevens1992MNRAS}, and favor a possible relation between black widows and isolated MSPs.

\section{SUMMARY}
Although the recycling scenario for MSPs have been well established, there remain some fundamental issues in the evolution of LMXBs, e.g., the mismatch between the birthrates of LMXBs and binary MSPs, the formation mechanism of isolated MSPs. The observations of black widow and redback MSPs demonstrate  remarkable feedback of the recycled NSs on the evolution of both the binary orbits and the secondary stars. We have calculated the LMXB evolution taking account of the possible effect of donor evaporation with different mass loss and AML modes, and found that evaporation can significantly alter the evolutionary paths and influence the properties of the final products. Our main conclusions are as follows.

1) For LMXBs with initial orbital period above the bifurcation period, the descendants are MSP/He WD binaries, but their distribution deviates from the standard $M_{\rm WD}-P_{\rm orb}$ relation for $P_{\rm orb}<20$ d. The WDs tend to be less massive for a given orbital period. This may explain some peculiar MSP systems like PSR J0751$+$1807.

2) LMXBs below the bifurcation period evolve along CV-like tracks, and evaporation-induced mass loss can significantly accelerate the evolution and result in dynamically unstable mass transfer, which may eventually leads to the disruption of the MSP companions and the formation of isolated MSPs.

\begin{acknowledgements}
We are grateful to an anonymous referee for helpful comments. This work was supported by the Natural Science Foundation of China under grant Nos. 11133001 and 11333004, the Strategic Priority Research Program of CAS under grant No. XDB09010200, and the Munich Institute for Astro- and Particle Physics (MIAPP) of the DFG cluster of excellence "Origin and Structure of the Universe".

\end{acknowledgements}



\newpage

\newpage

\newpage

\newpage

\newpage

\newpage

\newpage
\begin{figure}
\centering
\plotone{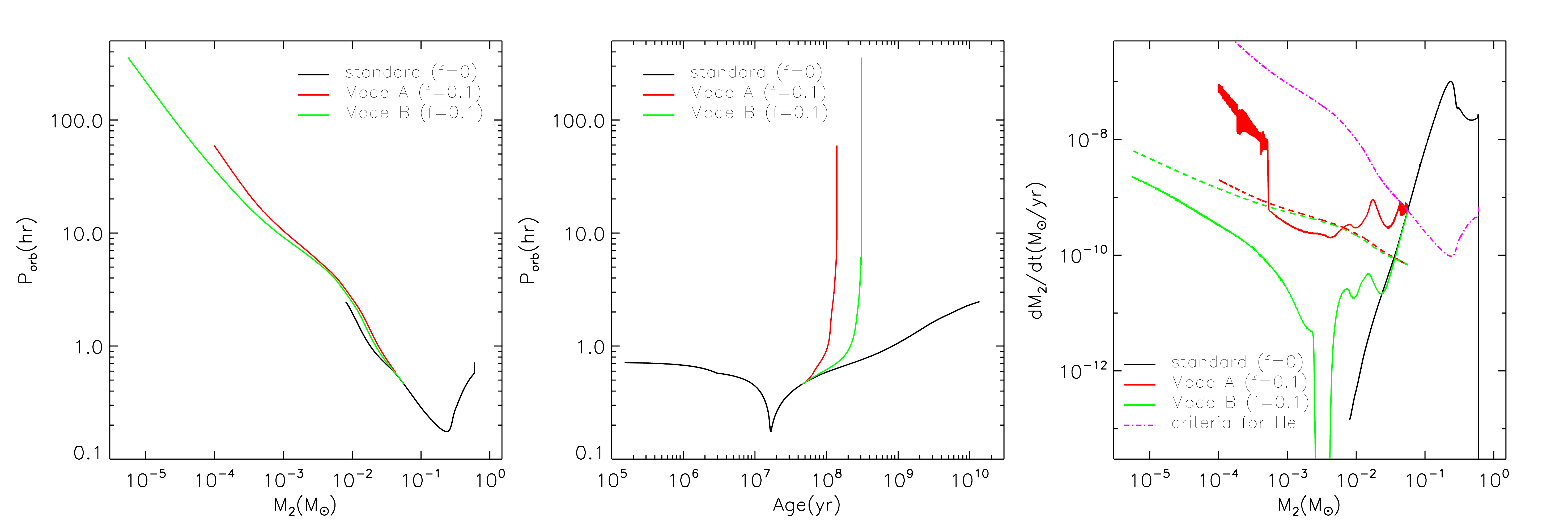}
\caption{Evolution of an UCXB with a 1.3 $M_{\odot}$ NS and a 0.6 $M_{\odot}$ He star in an initial 40 min orbit. The black line represents the standard evolution (without evaporation), and the red and green lines the evolutions with Modes A and B evaporation, respectively. The solid and dotted lines in the right panel represent the RLOF rate and evaporative mass loss rate, respectively. The  dashed magenta line denotes the critical mass transfer rate  for irradiated He disks \citep{Lasota2008AA}.
\label{figure7}}
\end{figure}

\end{document}